\documentclass[aps,prl,reprint,superscriptaddress,amsmath,amssymb]{revtex4-2}

\usepackage{graphicx}
\usepackage{txfonts}
\usepackage[hypertexnames=false]{hyperref}
\usepackage{braket}
\usepackage{bm}

\begin{document}

\title{Dynamical phase transitions in single particle Brownian motion without drift}
\author{Takahiro Kanazawa}
\affiliation{Department of Physics, The University of Tokyo, 7-3-1 Hongo, Bunkyo-ku, Tokyo 113-0033, Japan}
\affiliation{Nonequilibrium Physics of Living Matter RIKEN Hakubi Research Team, RIKEN Center for Biosystems Dynamics Research, 2-2-3 Minatojima-minamimachi, Chuo-ku, Kobe 650-0047, Japan}
\author{Kyogo Kawaguchi}
\affiliation{Department of Physics, The University of Tokyo, 7-3-1 Hongo, Bunkyo-ku, Tokyo 113-0033, Japan}
\affiliation{Nonequilibrium Physics of Living Matter RIKEN Hakubi Research Team, RIKEN Center for Biosystems Dynamics Research, 2-2-3 Minatojima-minamimachi, Chuo-ku, Kobe 650-0047, Japan}
\affiliation{RIKEN Cluster for Pioneering Research, 2-2-3 Minatojima-minamimachi, Chuo-ku, Kobe 650-0047, Japan}
\affiliation{Institute for Physics of Intelligence, The University of Tokyo, 7-3-1 Hongo, Bunkyo-ku, Tokyo 113-0033, Japan}
\author{Kyosuke Adachi}
\affiliation{Nonequilibrium Physics of Living Matter RIKEN Hakubi Research Team, RIKEN Center for Biosystems Dynamics Research, 2-2-3 Minatojima-minamimachi, Chuo-ku, Kobe 650-0047, Japan}
\affiliation{RIKEN Interdisciplinary Theoretical and Mathematical Sciences Program, 2-1 Hirosawa, Wako 351-0198, Japan}

\date{\today}

\begin{abstract}
Dynamical phase transitions (DPTs) arise from qualitative changes in the long-time behavior of stochastic trajectories, often observed in systems with kinetic constraints or driven out of equilibrium. 
Here we demonstrate that first-order DPTs can occur even in the large deviations of a single Brownian particle without drift, but only when the system's dimensionality exceeds four.
These DPTs are accompanied by temporal phase separations in the trajectories and exhibit dimension-dependent order due to the threshold behavior for bound state formation in Schr\"{o}dinger operators. 
We also discover second-order DPTs in one-dimensional Brownian motion, characterized by universal exponents in the rate function of dynamical observables.
Our results establish a novel framework linking classical DPTs to quantum phase transitions.
\end{abstract}

\maketitle

\textit{Introduction}.---A unique property inherent in atypical fluctuations in stochastic dynamics is the dynamical phase transition (DPT)~\cite{Touchette2009,garrahan2007dynamical}, a qualitative change in the dynamical path that appears upon changing the value of a time-integrated observable as a control parameter.
The order of DPT is determined by the singularity in the Legendre-Fenchel transform of the large deviation rate function, called the scaled cumulant generating function (SCGF)~\cite{Touchette2009}, which is a dynamical counterpart of the thermodynamic potential in the canonical ensemble~\cite{Chetrite2013}.
First-order DPTs have been found, for example, in kinetically constrained models~\cite{garrahan2007dynamical,garrahan2009first}, and second-order DPTs and related critical phenomena have been discussed in kinetic spin models~\cite{Jack2010}.

\begin{figure}[t]
    \centering
    \includegraphics[scale=1]{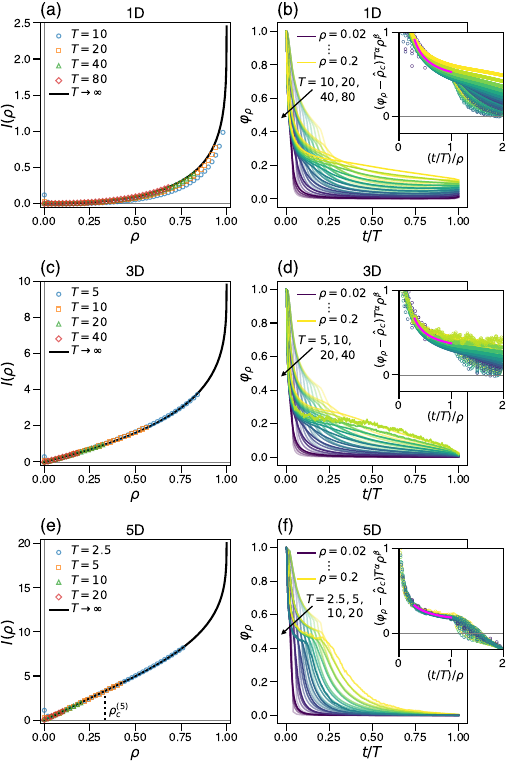}
    \caption{Dimensionality-induced phase separation.
    (a, c, e) Rate function $I(\rho)$ for one, three, and five dimensions.
    In each panel, the colored symbols represent the data from simulations with several values of observation time $T$, and the solid line is the theoretical prediction for $T \to \infty$.
    In (e), the predicted $I(\rho)$ is strictly linear for $0 < \rho < \rho_c^{(5)}$ $(= 1/3)$ due to phase separation in five dimensions; in contrast, no strictly linear region exists in (a) one or (c) three dimensions.
    (b, d, f) Time dependence of the order parameter $\varphi_\rho (t)$ for each dimension.
    We plot the data obtained from simulations with different values of $T$, where the brightness suggests the size of $\rho$.
    The inset in each panel is a scaling plot of $\varphi_\rho (t)$ using the two exponents $\alpha > 0$ and $\beta > 0$ and the nominal critical point $\hat{\rho}_c$, which are obtained by fitting within the plateau-like region (magenta line suggesting the fitting curve).
    We set the time step as $dt = 0.05$ and took $10^{12}$ independent samples for all simulations, and the bin size for $\rho$ was set to $0.02$.}
    \label{fig_dimensionality_ps}
\end{figure}

One of the simplest examples of DPT is the first-order DPT observed in a one-dimensional Brownian motion with drift~\cite{Nyawo2017,Nyawo2018}, which is related to the localization transition in non-Hermitian quantum models~\cite{Hatano1997}.
Intriguingly, a phase separation-like behavior in the temporal domain has been observed for the drifting Brownian motion; when focusing on trajectories with a certain occupation time, there is a typical time frame where the particle stays near the starting point before it wanders away~\cite{Nyawo2017} (see also Fig.~\ref{fig_dimensionality_ps}).
An interesting question is whether such DPT and phase separation-like behavior in the temporal domain can take place in an even simpler setting, i.e., a Brownian particle with no drift.

A candidate situation to observe DPTs in a system without drift and interactions will be at high dimensions. 
The eigenvalue problem at high dimensions has been considered in the context of high-energy physics~\cite{Nieto2002}.
Fluctuation at higher dimensions has been shown for example in quasi-crystals to affect quantities such as heat capacity in real three-dimensional materials~\cite{Yamamoto1996,Nagai2024}.
Another idea to observe DPTs is to explore distinct observables, which amounts to considering various interaction potentials between quantum particles that will lead to a bound state, which is crucial in many-body physics such as the BCS-BEC crossover~\cite{Randeria2014}.

In this Letter, we report on the two scenarios of DPTs appearing in the Brownian motion without drift.
First, we point out that a first-order DPT with temporal phase separation appears for the occupation fluctuations if the spatial dimension is higher than four.
Second, we find that a localization transition can appear even in one dimension as a second-order DPT when we observe the difference in occupations between two regions.
For the second case, we demonstrate that universal exponents appear in the rate function, independent of the detail of the observable.

\begin{table*}[t]
\centering
\begin{tabular}{c || c |  c |  c |  c |  c |  c}
Spatial dimension $d$ &  $1$  &  $2$ &  $3$  &  $4$  &  $5$ & $d$ ($\geq 5$) \\
\hline
SCGF $\lambda(k)$ for $\lambda(k) \approx 0$ &
$ k^2 $ & $ 4e^{-2\gamma}e^{-4/k} $ & $ \frac{1}{4} \left( k - \frac{\pi^2}{4} \right)^2 $ & $ - \frac{k - k_c^{(4)}}{\ln{\left( k - k_c^{(4)} \right)}} $ & $ \frac{1}{3} (k - \pi^2) $ & $ \frac{d-4}{d-2} (k - k_c^{(d)}) $ \\
\hline
Rate function $I(\rho)$ for $\rho \approx 0$ & $\approx \frac{1}{4} \rho^2$ & $\approx -4 \frac{\rho}{\ln{\rho}}$ & $\approx \frac{\pi^2}{4} \rho$ & $\approx k_c^{(4)} \rho$ & $=\pi^2 \rho$ $(0 < \rho < \frac{1}{3})$ & $= k_c^{(d)} \rho$ $(0 < \rho < \rho_c^{(d)} = \frac{d-4}{d-2})$ \\
\hline
Order of DPT & $2^*$ & - & $2$ & $2$ & $1$ & $1$ \\
\end{tabular}
\caption{Leading order behavior of the SCGF $\lambda(k)$ and the rate function $I(\rho)$ and the orders of DPT for different spatial dimensions $d$.
When $d$ is five or higher, $\lambda(k)$ is undifferentiable at $k=k_c^{(d)}$, suggesting a first-order DPT; correspondingly, $I(\rho)$ is a strictly linear function for a finite domain $0 < \rho < \rho_c^{(d)}$, where the temporal phase separation is expected to appear.
As $d \to \infty$, we asymptotically obtain $k_c^{(d)} \approx d^2/4$ and $\rho_c^{(d)} \approx 1 - 2/d$.
Regarding the second order DPTs, the second derivative of $\lambda (k)$ is discontinuous at $k = 0$ for $d = 1$ (*), which is distinct from $d = 3$ and $d = 4$ where $k = k_c^{(d)}$ ($> 0$).
See Ref.~\cite{Full_arxiv} for the derivations.}
\label{tab_expanded_scgf}
\end{table*}

\textit{Dimensionality-induced temporal phase separation}.---We consider the dynamics of a Brownian particle in $d$ dimensions:
\begin{equation}
    \frac{d\bm{x}(t)}{dt} = \bm{\xi}(t),
    \label{eq:SDE}
\end{equation}
where $\bm{\xi}(t)$ is a Gaussian noise with $\braket{\xi_a(t)} = 0$ and $\braket{\xi_a(t) \xi_b(t')} = 2 \delta_{ab}\delta(t-t')$ ($a,b \in \{ 1,2,\cdots,d \}$).
Here, the diffusion constant is set to unity by rescaling the time, and the initial condition is $\bm{x}(0) = \bm{0}$.
As a time-averaged observable, we take the fraction of time spent by the particle within the $d$-dimensional unit ball centered at the origin:
\begin{equation}
    \rho_T := \frac{1}{T} \int^T_0 \chi_{[0,1]}\bm{(}r(t)\bm{)} dt.
    \label{eq_observable}
\end{equation}
Here, $r(t) := ||\bm{x}(t)||$, $T$ is the total observation time, and $\chi_A(z)$ is the indicator function, which takes $\chi_A(z) = 1$ for $z \in A$ and $\chi_A(z) = 0$ otherwise.
By definition, $0 < \rho_T \leq 1$.

As $T$ becomes large, the probability of $\rho_T$ taking a non-zero value becomes vanishingly small. According to the large deviation principle~\cite{Touchette2009}, the probability density of $\rho_T$ at $T \to \infty$ is:
\begin{equation}
    P(\rho_T=\rho) = e^{-TI(\rho) + o(T)},
\end{equation}
where $I(\rho)$ is the rate function, which becomes singular upon the appearance of DPT.
We can obtain $I(\rho)$ from the SCGF,
\begin{equation}
    \lambda(k) := \lim_{T \to \infty} \frac{1}{T} \ln{\langle e^{Tk\rho_T} \rangle},
\end{equation}
by the Legendre-Fenchel transformation: $I(\rho) = \max_k \{k \rho - \lambda (k)\}$, if $\lambda(k)$ is differentiable~\cite{Touchette2009}.
From Eqs.~\eqref{eq:SDE} and \eqref{eq_observable}, we can show that $\lambda(k)$ is the dominant eigenvalue of the biased generator~\cite{Touchette2018} $\mathcal{L}_k = \bm{\nabla}^2 + k\chi_{[0, 1]} (r)$, where the corresponding eigenfunction $\phi_k (\bm{x})$ satisfies
\begin{equation}
    \mathcal{L}_k \phi_k (\bm{x}) = \lambda (k) \phi_k (\bm{x}).
    \label{eq_spectrum}
\end{equation}
Note that $-\mathcal{L}_k$ is equivalent to the quantum Hamiltonian of a particle under a potential well with depth $k$, and $\phi_k (\bm{x})$ corresponds to the ground state wavefunction.

If $\lambda(k)$ is differentiable and the condition for the ensemble equivalence is satisfied~\cite{Chetrite2015,Agranov2020}, the probability density for the particle position $\bm{x}$ conditioned with $\rho_T = \rho$, $P(\bm{x} | \rho_T = \rho)$, is proportional to $\phi_{k^*(\rho)}(\bm{x})^2$, apart from the time domain near $t = 0$ or $t = T$.
Here, $k^*(\rho) := \mathrm{argmax}_k \{k\rho - \lambda(k)\}$, so $\phi_{k^*(\rho)}(\bm{x})^2$ corresponds to the ground state wavefunction calculated for an appropriate amplitude $k^*(\rho)$ set for the potential energy in the quantum problem.
In particular, when $\phi_{k^*(\rho)} (\bm{x})$ is a wavefunction localized at some region, the conditional distribution $P(\bm{x} | \rho_T = \rho)$ should also be localized around the same region.

Since the dominant eigenfunction $\phi_k (\bm{x})$ depends only on $r = ||\bm{x}||$ [i.e., $\phi_k(\bm{x}) = \phi_k(r)$] due to the rotational symmetry of $\mathcal{L}_k$, the eigenvalue equation [Eq.~\eqref{eq_spectrum}] reduces to
\begin{equation}
    \left[ \frac{d^2}{dr^2} + \frac{d-1}{r} \frac{d}{dr} + k\chi_{[0,1]}(r) \right] \phi_k(r) = \lambda(k) \phi_k(r).
    \label{eq_reduced_eigeq}
\end{equation}
By solving Eq.~\eqref{eq_reduced_eigeq}~\cite{Nieto2002}, we obtain $ \phi_k(r) = A_k f (r \sqrt{k-\lambda(k)}; d) \theta (1-r) + B_k g (r \sqrt{\lambda(k)}; d) \theta (r-1)$, where $f(z; d):=J_{(d-2)/2}(z)/z^{(d-2)/2}$ and $g(z; d):=K_{(d-2)/2}(z)/z^{(d-2)/2}$ with $J_{\nu} (z)$ and $K_{\nu} (z)$ being the Bessel functions of the first kind and the modified Bessel functions of the second kind with order $\nu$, respectively, and $\theta(z)$ is the Heaviside step function~\cite{Full_arxiv}.
The ratio $A_k/B_k$ and the eigenvalue $\lambda(k)$ are determined by the continuity conditions of $\phi_k(r)$ and $d\phi_k(r) / dr$ at $r = 1$~\cite{Full_arxiv}.

For $\lambda(k) \approx 0$, we can analytically obtain the functional form of $\lambda(k)$~\cite{Full_arxiv}.
As known in the corresponding quantum problem for the ground state~\cite{Landau_quantum}, when $d = 1$ or $d = 2$, $\lambda (k)$ has no singularity for $k \geq 0$, and $\lambda (k) > 0$ for $k > 0$.
When $d = 3$~\cite{Sahu1989,Full_arxiv}, $\lambda (k)$ smoothly changes from zero (for $k \leq k_c^{(3)} := \pi^2/4$) to positive as $\lambda(k) \propto (k - k_c^{(3)})^2$ (for $k > k_c^{(3)}$), suggesting a second-order DPT at $k = k_c^{(3)}$.
Here, a positive $\lambda (k)$ suggests that $\phi_k(r)$ is localized around the well (i.e., $r \in [0, 1]$) and a bound state is formed in the quantum counterpart.
This type of DPT has been predicted in Ref.~\cite{Nyawo2018}.
When $d = 4$, similarly to the case of $d = 3$, $\lambda (k)$ smoothly becomes positive at $k = k_c^{(4)}$ ($> 0$)~\cite{Full_arxiv}.
In contrast, when $d \geq 5$, $\lambda (k)$ changes from zero (for $k \leq k_c^{(d)}$) to positive as $\lambda (k) \propto k - k_c^{(d)}$ (for $k > k_c^{(d)}$)~\cite{Full_arxiv}, suggesting a first-order DPT at a dimension-dependent critical value $k = k_c^{(d)}$ ($>0$).
When $d \to \infty$, the positive constant of proportionality approaches $1$ while $k_c^{(d)} \to \infty$ [i.e., $\lambda (k) \approx (1-2/d) (k-d^2/4)$]~\cite{Full_arxiv}.
The behavior of $\lambda(k)$ and $I(\rho)$ for each dimension as well as the orders of DPT are summarized in Table~\ref{tab_expanded_scgf}.

The DPT seen in $\lambda(k)$ can lead to a singularity in the rate function $I(\rho)$ through the Legendre-Fenchel transformation.
The smooth $\lambda(k)$ in four or lower dimensions only results in a smooth $I(\rho)$ for any $\rho > 0$, where no qualitative change in the dynamical path is expected.
In contrast, the first-order DPT in five or higher dimensions suggests the existence of a strictly linear region of $I(\rho)$ [i.e., $I(\rho) \propto \rho$] for $0 < \rho < \rho_c^{(d)}$ with a dimension-dependent $\rho_c^{(d)} := (d-4)/(d-2)$~\cite{Full_arxiv}.
Specifically, for $d = 5$, we find that $I(\rho) = \pi^2 \rho$ for $\rho$ smaller than $\rho_c^{(5)} = 1/3$, and $\rho_c^{(d)} \to 1$ for $d \to \infty$ as we show in Ref.~\cite{Full_arxiv}.

It has been previously indicated~\cite{Nyawo2017,Nyawo2018} that the existence of a linear part in $I(\rho)$ leads to temporal phase separation of the dynamical path: a particle trajectory separates into two segments, the first localized around the origin, and the second being non-localized.
This phenomenon is similar to the spatial phase separation in thermodynamics, where the Helmholtz free energy linearly depends on the particle density~\cite{Landau_statphys}.
To see this, we performed simulations of Brownian motion at high dimensions ($d = 5$), compared with lower dimensions ($d = 1$ and $3$).
In Figs.~\ref{fig_dimensionality_ps}(a), (c), and (e), we plot $I(\rho)$ obtained from simulations for $d = 1$, $3$, and $5$, respectively, which agree with the theoretical predictions.
In particular, for $d = 5$ with $0 < \rho < \rho_c^{(5)}$, the predicted linear dependence of $I(\rho)$ is asymptotically reproduced as $T$ increases.

To see the phase separation, we define a time-dependent order parameter:
\begin{equation}
    \varphi_\rho (t) := \braket{\chi_{[0,1]}\bm{(}r(t)\bm{)}}_{\rho},
\end{equation}
which probes the localization in the $d$-dimensional unit ball centered at the origin at time $t$, conditioned with $\rho_T = \rho$.
Here, $\braket{\cdots}_\rho$ denotes the expectation value with respect to the conditional probability $P(\bm{x}|\rho_T=\rho)$.
When the phase separation appears as $T \to \infty$ in five or higher dimensions, we expect temporally separated two phases: the localized phase with $\varphi_\rho(t) = \rho_c^{(d)}$ for the time domain $t/T \in (0, \rho/\rho_c^{(d)})$ and the non-localized phase with $\varphi_\rho(t) = 0$ for the remaining time domain $t/T \in (\rho/\rho_c^{(d)}, 1)$.
As a key feature of phase separation, $\varphi_\rho (t)$ in each phase does not depend on the specific value of $t/T$ or $\rho$ when $T$ is large enough.
Here, the value of $\varphi_\rho(t)$ and the size of the time domain in each phase are derived from the lever rule that generally holds in spatial phase separation~\cite{Rubinstein2003} and is expected to hold in temporal phase separation~\cite{Nyawo2018}.
The initial condition [$r(0) = 0$] sets the localized phase to appear in the first time domain [i.e., $t/T \in (0, \rho/\rho_c^{(d)})$].
Note that $T^{-1} \int_0^T dt \varphi_\rho(t) = \rho$ for any $\rho \in (0, \rho_c^{(d)})$, which is consistent with the condition $\rho_T = \rho$.
For $d \to \infty$, as $\rho_c^{(d)} \to 1$, we expect completely separated phases either with $\varphi_{\rho}(t) = 1$ or $\varphi_{\rho}(t) = 0$.

In Figs.~\ref{fig_dimensionality_ps}(b), (d), and (f), we plot the time dependence of $\varphi_\rho(t)$ with several values of $\rho$ and $T$, obtained from simulations for $d = 1$, $3$, and $5$, respectively.
For $d = 5$, $\varphi_\rho(t)$ shows a plateau-like behavior with a $\rho$-independent height as $T$ increases.
According to the property of phase separation, the height of this plateau-like region should approach $\rho_c^{(5)}$ ($= 1/3$) as $T \to \infty$, independently of $t/T$ or $\rho$.
To confirm this, we looked for an asymptotic scaling law for $T \to \infty$ by fitting $\varphi_\rho (t)$ within the plateau-like region.
Assuming a power-law scaling for $T$ and $\rho$, we fitted the observed $\varphi_\rho (t)$ by a function $F(t, T, \rho) := \hat{\rho}_c + T^{-\alpha} \rho^{-\beta} f\bm{(}(t/T)/\rho\bm{)}$ with $f(x) := (a_0 + a_1 x) / (1 + b_1 x)$, where $\alpha$, $\beta$, $\hat{\rho}_c$, $a_0$, $a_1$, and $b_1$ are fitting parameters~\cite{Full_arxiv}.
The scaling plots with the optimal parameters for $d = 1$, $3$, and $5$ are shown in the insets of Figs.~\ref{fig_dimensionality_ps}(b), (d), and (f), respectively, where the magenta lines represent the best-fitted functions.
We find a clear scaling around the plateau-like region for $d = 5$ (with $\alpha \approx 0.50$, $\beta \approx 0.46$, and $\hat{\rho}_c \approx 0.32$), suggesting that $\varphi_\rho - \hat{\rho}_c \sim T^{-\alpha} \to 0$, i.e., $\varphi_\rho \to \hat{\rho}_c$ ($\approx \rho_c^{(5)}$) for $T \to \infty$, consistent with the predicted phase separation.
In contrast, we do not see a similar clear scaling for $d = 1$ or $3$, suggesting that there is no phase separation, as expected.

\begin{figure}[t]
    \centering
    \includegraphics[scale=1]{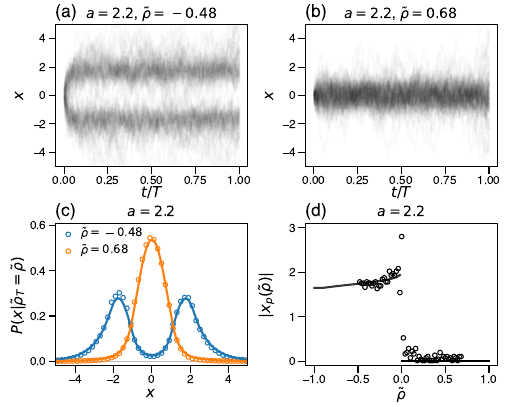}
    \caption{Localization transition as DPTs in one-dimensional Brownian motion (same as Fig.~8 in Ref.~\cite{Full_arxiv}).
    (a, b) 100 typical trajectories of the particle conditioned with (a) $\tilde{\rho}=-0.48$ and (b) $\tilde{\rho}=0.68$.
    (c) Conditional probability density $P(x|\tilde{\rho}_T=\tilde{\rho})$ for $\tilde{\rho}=-0.48$ and $\tilde{\rho}=0.68$, corresponding to (a) and (b), respectively.
    The colored circles and solid lines are obtained from simulations and the analytical expression, i.e., $\phi_{k^*(\tilde{\rho})}(x)^2$ (see Ref.~\cite{Full_arxiv} for the functional form), respectively, showing good agreement.
    The numerical $P(x|\tilde{\rho}_T=\tilde{\rho})$ is calculated from $10^3$ independent trajectories in the time domain of $1/4<t/T<3/4$ with the bin size of $\tilde{\rho}$ being $0.02$.
    (d) $\tilde{\rho}$ dependence of the peak position of $P(x|\tilde{\rho}_T=\tilde{\rho})$, $|x_p(\tilde{\rho})|$, which shows a jump at $\tilde{\rho}=0$.
    The circles and solid lines are obtained from simulations and the analytical expression (see Ref.~\cite{Full_arxiv}), respectively.
    The numerical $|x_p(\tilde{\rho})|$ is calculated from $10^3$ independent trajectories in the time domain of $1/4<t/T<3/4$ for each $\tilde{\rho}$ with the bin sizes of $\tilde{\rho}$ and $|x|$ being $0.02$ and $0.03$, respectively.
    Parameters for (a-d): $a=2.2$, 
    $dt=0.05$, and $T=20$.}
    \label{fig2_22}
\end{figure}

\textit{Localization transition in one dimension}.---The result in the previous section suggests that no DPT appears in the trajectories of one-dimensional Brownian motion as long as we take the simple observable defined in the form of Eq.~\eqref{eq_observable}.
Here, we search for possible DPTs in one dimension by modifying the observable; we consider the difference between the fraction of time satisfying $|x(t)| \in [0, 1]$ (length scale rescaled so that the upper bound is unity) and the fraction of time satisfying $|x(t)| \in [1, a]$:
\begin{equation}
    \tilde{\rho}_T := \frac{1}{T} \int^T_0 \chi_{[0,1]}\bm{(}|x(t)|\bm{)} dt - \frac{1}{T} \int^T_0 \chi_{[1,a]}\bm{(}|x(t)|\bm{)} dt.
    \label{eq_observable_diff}
\end{equation}
We have $|x(t)| = r(t)$ in one dimension, and we set $x(0) = 0$ as the initial condition.
Replacing $\rho_T$ by $\tilde{\rho}_T$ and $\rho$ by $\tilde{\rho}$, we can use the same formulation as explained in the previous section; for example, the SCGF is defined as $\lambda(k) := \lim_{T \to \infty} T^{-1} \ln \braket{e^{T k \tilde{\rho}_T}}$.

When the time fraction for $|x(t)| \in [1, a]$ is larger or smaller than that for $|x(t)| \in [0, 1]$, $\tilde{\rho}_T$ takes a negative or positive value, respectively.
Thus, we can expect that the typical trajectories conditioned with $\tilde{\rho}_T = \tilde{\rho}$ will switch from those localized around $|x(t)| \in [1, a]$ for $\tilde{\rho} < 0$ to those localized around $|x(t)| \in [0, 1]$ for $\tilde{\rho} > 0$.
In the following, we analytically and numerically show that this localization transition indeed appears, which corresponds to second-order DPTs in $\lambda(k)$.

In Figs.~\ref{fig2_22}(a) and (b), we plot numerically sampled Brownian particle trajectories that satisfy  $\tilde{\rho}_{T} = \tilde{\rho}$ with $\tilde{\rho} = -0.48$ and $\tilde{\rho} = 0.68$, respectively, with the parameter $a = 2.2$.
As expected, the trajectories are localized around $|x(t)| \in [1, a] = [1, 2.2]$ for $\tilde{\rho} = -0.48$ ($< 0$), while localized around $|x(t)| \in [0, 1]$ for $\tilde{\rho} = 0.68$ ($> 0$).
In Fig.~\ref{fig2_22}(c), we plot the empirical distribution of the particle position (colored circles), $P(x|\tilde{\rho}_T=\tilde{\rho})$, for $\tilde{\rho} = -0.48$ and $\tilde{\rho} = 0.68$ [corresponding to Figs.~\ref{fig2_22}(a) and (b), respectively].
We find clear agreement with the analytically obtained asymptotic distribution for $T \to \infty$ (colored lines, see Ref.~\cite{Full_arxiv} for the derivation).

Figure~\ref{fig2_22}(c) suggests that the peak position(s) of $P(x|\tilde{\rho}_T=\tilde{\rho})$, denoted by $x_p(\tilde{\rho}) := \mathrm{argmax}_x\{P(x|\tilde{\rho}_T=\tilde{\rho})\}$, is distinct between the two regimes of $\tilde{\rho}$; $|x_p(\tilde{\rho})|>0$ for $\tilde{\rho}<0$ and $|x_p(\tilde{\rho})|=0$ for $\tilde{\rho}>0$.
As shown with black circles in Fig.~\ref{fig2_22}(d), we numerically confirm this localization transition as the jump of $|x_p(\tilde{\rho})|$ at $\tilde{\rho} = 0$, which is also supported by the analytical expression of $|x_p(\tilde{\rho})|$ for $T \to \infty$~\cite{Full_arxiv} (black lines).

\begin{figure}[t]
    \centering
    \includegraphics[scale=1]{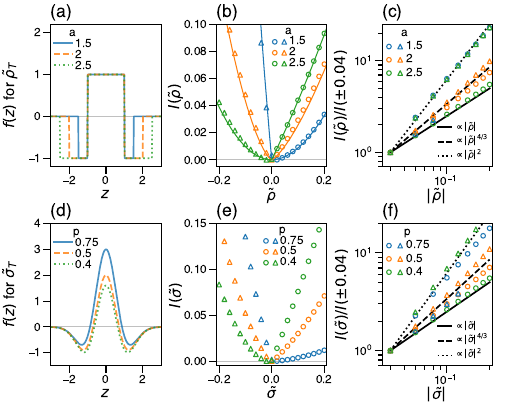}
    \caption{Universal exponents of the rate functions in one-dimensional Brownian motion (same as Fig.~7 in Ref.~\cite{Full_arxiv}).
    (a) Integrand of the observable $\tilde{\rho}_T$, $f(z)=\chi_{[0,1]}(|z|)-\chi_{[1,a]}(|z|)$, with $a=1.5$ (blue solid line), $2.0$ (orange dashed line), and $2.5$ (green dotted line).
    These parameters correspond to $\int_{-\infty}^{\infty} f(z)dz > 0$, $\int_{-\infty}^{\infty} f(z)dz = 0$, and $\int_{-\infty}^{\infty} f(z)dz < 0$, respectively.
    (b) Rate functions of $\tilde{\rho}$, whose integrands are shown in (a).
    The colored symbols are the simulation results, and the solid lines are theoretical predictions.
    The data show non-differentiability at $\tilde{\rho} = 0$ for $a = 1.5$, and $2.5$.
    Parameters are set to $dt = 0.05$, $T = 80$ for simulations and we take $10^8$ samples to calculate the rate function.
    The number of trajectories are counted with bin size of $\tilde{\rho}$ being $0.02$ and normalized so that $I(\tilde{\rho})$ takes $0$ at its minimum.
    (c) Log-log plot of (b).
    Open circles and triangles are the simulation results for $\tilde{\rho}>0$ and $\tilde{\rho}<0$, respectively.
    The black lines are power fuctions with the theoretically predicted exponents.
    (d) Integrand of the observable $\tilde{\sigma}_T$, $f(z) = -4e^{-z^2}(z^2-p)$.
    We plot for $p=0.75$ (blue solid line), $0.50$ (orange dashed line), and $0.40$ (green dotted line).
    These parameters correspond to $\int_{-\infty}^{\infty} f(z)dz > 0$, $\int_{-\infty}^{\infty} f(z)dz = 0$, and $\int_{-\infty}^{\infty} f(z)dz < 0$, respectively.
    (e) Rate functions of $\tilde{\sigma}$, whose integrands are shown in (d).
    Simulation conditions are the same as (b).
    (f) Log-log plot of (e).
    Open circles and triangles are the simulation results for $\tilde{\sigma} > 0$ and $\tilde{\sigma} < 0$, respectively.}
    \label{fig3_23}
\end{figure}

\begin{table*}[t]
\centering
\begin{tabular}{c || c |  c |  c |  c |  c}
& \multicolumn{2}{|c|}{$\int_{-\infty}^{\infty} f(z) dz > 0$}  & $\int_{-\infty}^{\infty} f(z) dz = 0$ & \multicolumn{2}{|c}{$\int_{-\infty}^{\infty} f(z) dz < 0$} \\
\hline
SCGF $\lambda(k)$ for $\lambda(k) \approx 0$ & $ k^2$ & $(k-k_c)^2$ & $k^4$ & $(k-k_c)^2$ & $k^2$ \\
 & $( k > 0,\ k \approx 0)$ & $(k < k_c<0,\ k \approx k_c)$ & $(k \approx 0)$ & $(k > k_c>0,\ k \approx k_c)$ & $(k < 0,\ k \approx 0)$ \\
\hline
Rate function $I(\sigma)$ for $\sigma \approx 0$ & $\sigma^2$ & $|\sigma|$ & $|\sigma|^{4/3}$ & $\sigma$ & $\sigma^2$  \\
 & $( \sigma > 0,\ \sigma \approx 0)$ & $(\sigma<0,\ \sigma \approx 0)$ & $(\sigma \approx 0)$ & $(\sigma>0,\ \sigma \approx 0)$ & $(\sigma < 0,\ \sigma \approx 0)$ \\
\end{tabular}
\caption{Power-law asymptotic forms of the SCGF $\lambda(k)$ and the rate function $I(\sigma)$ for three cases of the observable integrand $f(z)$ in one-dimensional Brownian motion.
We omit the constants of proportionality.
See Ref.~\cite{Full_arxiv} for the derivations.}
\label{tab_expanded_scgf_1d}
\end{table*}

\textit{Universal properties of localization transition}.---The localization transition found in the previous section should lead to a singular behavior of the rate function $I(\tilde{\rho})$ at $\tilde{\rho} = 0$.
In Fig.~\ref{fig3_23}(b), we show $I(\tilde{\rho})$ obtained from simulations (colored symbols) for three parameters $a = 1.5$, $2.0$, and $2.5$ [see Fig.~\ref{fig3_23}(a) for the functional form of the integrand of $\tilde{\rho}_T$ in Eq.~\eqref{eq_observable_diff}].
For $a = 1.5$ and $2.5$, $I(\tilde{\rho})$ becomes non-differentiable at $\tilde{\rho} = 0$, which is clearly seen in the log-log plot of $I(\tilde{\rho})$ vs $|\tilde{\rho}|$ [Fig.~\ref{fig3_23}(c)].
Furthermore, we find a power-law behavior of $I(\tilde{\rho})$ for $\tilde{\rho} \approx 0$: $I(\tilde{\rho}) \approx |\tilde{\rho}|$, $I(\tilde{\rho}) \approx |\tilde{\rho}|^{4/3}$, or $I(\tilde{\rho}) \approx \tilde{\rho}^2$, depending on the sign of $\tilde{\rho}$ and the value of $a$ [see Fig.~\ref{fig3_23}(c)].

Solving the eigenvalue problem for the current setup, i.e., Eq.~\eqref{eq_spectrum} with $\mathcal{L}_k = d^2/dx^2 + k[\chi_{[0,1]}(|x|) - \chi_{[1,a]}(|x|)]$, we can obtain the SCGF $\lambda(k)$ and the rate function $I(\tilde{\rho})$ for $T \to \infty$, in a similar way described in the previous section~\cite{Full_arxiv}.
The obtained $I(\tilde{\rho})$ for $T \to \infty$ is plotted as colored solid lines for each parameter set in Fig.~\ref{fig3_23}(b), showing quite good agreement with the data from simulations with $T = 80$ (colored symbols).

The non-differentiability in $I(\tilde{\rho})$ indicates that the SCGF $\lambda(k)$ also shows a singular behavior.
The result obtained in Ref.~\cite{Full_arxiv} suggests that, for $a \neq 2$, second-order DPTs appear at two transition points upon changing $k$ from $-\infty$ to $\infty$; the localized state around $|x| \in [1, a]$ first changes into the non-localized state, and then changes into the other localized state around $|x| \in [0, 1]$.
In contrast, for $a = 2$, where the two transition points merge at $k = 0$, the SCGF becomes quartic as $\lambda(k) \propto k^4$, which corresponds to $I(\tilde{\rho}) \propto |\tilde{\rho}|^{4/3}$.

The power law of $I(\tilde{\rho})$ and the corresponding second-order DPTs in $\lambda(k)$ remind us of the universal critical phenomena in phase transitions~\cite{Hohenberg1977,Chaikin1995}.
To consider the possible universality of the power-law exponents of the rate function, we consider a general time-averaged observable:
\begin{equation}
    \sigma_T := \frac{1}{T} \int^T_0 f\bm{(}x(t)\bm{)} dt,
\end{equation}
where $f(z)$ is a real scalar function with $\int^{\infty}_{-\infty} (1+|z|)|f(z)| dz<\infty$~\cite{Klaus1977}.

Here, we consider the asymptotic power-law behavior of the rate function $I(\sigma)$ for small $|\sigma|$.
We define a threshold $k_c$, such that $\lambda \to + 0$ when $k \to k_c + 0$, as in the previous section.
For $d=1$, the leading order of $\lambda(k)$ is shown to be proportional to $(k-k_c)^2$ when $\int^{\infty}_{-\infty} f(z)dz \neq 0$, while $(k-k_c)^4$ when the integral is $0$~\cite{Klaus1980,Simon1976,Klaus1977}.
Moreover, $k_c=0$ when $-\mathrm{sgn}(k) \int^{\infty}_{-\infty} f(z)dz \leq 0$, with $\mathrm{sgn}(z)$ being the sign function, otherwise $k_c \neq 0$.
By the Legendre-Fenchel transformation (see Ref.~\cite{Full_arxiv}), calculation of $I(\sigma)$ yields the leading order proportional to either $|\sigma|$, $|\sigma|^{4/3}$, or $\sigma^2$ depending on $f(z)$ (see Table~\ref{tab_expanded_scgf_1d}).
As another example to confirm the universality, we perform Brownian particle simulations with $f(z)=-4e^{-z^2}(z^2-p)$, where $p$ is a constant [see Fig.~\ref{fig3_23}(d) for the functinal forms].
We plot its rate function $I(\tilde{\sigma})$ in Fig.~\ref{fig3_23}(e), which shows non-differentiability at $\tilde{\sigma} = 0$ for $p = 0.75$ and $0.40$.
Figure~\ref{fig3_23}(f) is a log-log plot of Fig.~\ref{fig3_23}(e) and for small $|\tilde{\sigma}|$, the universal exponents $|\tilde{\sigma}|$, $|\tilde{\sigma}|^{4/3}$, and $\tilde{\sigma}^2$ are observed.

\textit{Discussion and conclusion}.---In this Letter, we proposed two types of dynamical phase transitions in the simple Brownian motion: the dimensionality-induced transitions and the observable-dependent transition in one dimension.
Although both of these predictions are testable in principle by experiments involving colloids, it is more likely that they will find applications outside of real Browninan particles, such as in finance~\cite{Bouchaud2003} and in machine learning models~\cite{Sohl2015} where multi-dimensional stochastic dynamics is naturally considered, or in many-body models in low dimensional quantum mechanics.
For example, from the general coupling constant threshold behavior of the SCGF shown in a context of spectrum theory~\cite{Klaus1980}, the first-order DPT is expected to be found in general time-averaged scalar observables.
We show in Ref.~\cite{Full_arxiv} that the time fraction that \emph{all} $N$ non-interacting Brownian particles in $d$-dimensions spend in some interval, for example, $r_i(t) \in [0,1]$ with $i=1,2, \dots, N$, will yield the same type of transition as far as $Nd \geq 5$.
Through the expanded cases of systems exhibiting DPTs, we expect not only a stronger bridge between classical dynamics and quantum mechanics but also the emergence of further application cases for DPTs across various fields.

\begin{acknowledgments}
We thank Takahiro Nemoto and Akira Shimizu for fruitful discussions.
This work was supported by JSPS KAKENHI Grant Numbers JP20K14435 (to K.A.), JP19H05795, JP19H05275, JP21H01007, and JP23H00095 (to K.K.).
\end{acknowledgments}


\begin{thebibliography}{27}%
\makeatletter
\providecommand \@ifxundefined [1]{%
 \@ifx{#1\undefined}
}%
\providecommand \@ifnum [1]{%
 \ifnum #1\expandafter \@firstoftwo
 \else \expandafter \@secondoftwo
 \fi
}%
\providecommand \@ifx [1]{%
 \ifx #1\expandafter \@firstoftwo
 \else \expandafter \@secondoftwo
 \fi
}%
\providecommand \natexlab [1]{#1}%
\providecommand \enquote  [1]{``#1''}%
\providecommand \bibnamefont  [1]{#1}%
\providecommand \bibfnamefont [1]{#1}%
\providecommand \citenamefont [1]{#1}%
\providecommand \href@noop [0]{\@secondoftwo}%
\providecommand \href [0]{\begingroup \@sanitize@url \@href}%
\providecommand \@href[1]{\@@startlink{#1}\@@href}%
\providecommand \@@href[1]{\endgroup#1\@@endlink}%
\providecommand \@sanitize@url [0]{\catcode `\\12\catcode `\$12\catcode `\&12\catcode `\#12\catcode `\^12\catcode `\_12\catcode `\%12\relax}%
\providecommand \@@startlink[1]{}%
\providecommand \@@endlink[0]{}%
\providecommand \url  [0]{\begingroup\@sanitize@url \@url }%
\providecommand \@url [1]{\endgroup\@href {#1}{\urlprefix }}%
\providecommand \urlprefix  [0]{URL }%
\providecommand \Eprint [0]{\href }%
\providecommand \doibase [0]{https://doi.org/}%
\providecommand \selectlanguage [0]{\@gobble}%
\providecommand \bibinfo  [0]{\@secondoftwo}%
\providecommand \bibfield  [0]{\@secondoftwo}%
\providecommand \translation [1]{[#1]}%
\providecommand \BibitemOpen [0]{}%
\providecommand \bibitemStop [0]{}%
\providecommand \bibitemNoStop [0]{.\EOS\space}%
\providecommand \EOS [0]{\spacefactor3000\relax}%
\providecommand \BibitemShut  [1]{\csname bibitem#1\endcsname}%
\let\auto@bib@innerbib\@empty
\bibitem [{\citenamefont {Touchette}(2009)}]{Touchette2009}%
  \BibitemOpen
  \bibfield  {author} {\bibinfo {author} {\bibfnamefont {H.}~\bibnamefont {Touchette}},\ }\bibfield  {title} {\bibinfo {title} {The large deviation approach to statistical mechanics},\ }\href {https://doi.org/10.1016/j.physrep.2009.05.002} {\bibfield  {journal} {\bibinfo  {journal} {Phys. Rep.}\ }\textbf {\bibinfo {volume} {478}},\ \bibinfo {pages} {1} (\bibinfo {year} {2009})}\BibitemShut {NoStop}%
\bibitem [{\citenamefont {Garrahan}\ \emph {et~al.}(2007)\citenamefont {Garrahan}, \citenamefont {Jack}, \citenamefont {Lecomte}, \citenamefont {Pitard}, \citenamefont {van Duijvendijk},\ and\ \citenamefont {van Wijland}}]{garrahan2007dynamical}%
  \BibitemOpen
  \bibfield  {author} {\bibinfo {author} {\bibfnamefont {J.~P.}\ \bibnamefont {Garrahan}}, \bibinfo {author} {\bibfnamefont {R.~L.}\ \bibnamefont {Jack}}, \bibinfo {author} {\bibfnamefont {V.}~\bibnamefont {Lecomte}}, \bibinfo {author} {\bibfnamefont {E.}~\bibnamefont {Pitard}}, \bibinfo {author} {\bibfnamefont {K.}~\bibnamefont {van Duijvendijk}},\ and\ \bibinfo {author} {\bibfnamefont {F.}~\bibnamefont {van Wijland}},\ }\bibfield  {title} {\bibinfo {title} {Dynamical first-order phase transition in kinetically constrained models of glasses},\ }\href {https://doi.org/10.1103/PhysRevLett.98.195702} {\bibfield  {journal} {\bibinfo  {journal} {Phys. Rev. Lett.}\ }\textbf {\bibinfo {volume} {98}},\ \bibinfo {pages} {195702} (\bibinfo {year} {2007})}\BibitemShut {NoStop}%
\bibitem [{\citenamefont {Chetrite}\ and\ \citenamefont {Touchette}(2013)}]{Chetrite2013}%
  \BibitemOpen
  \bibfield  {author} {\bibinfo {author} {\bibfnamefont {R.}~\bibnamefont {Chetrite}}\ and\ \bibinfo {author} {\bibfnamefont {H.}~\bibnamefont {Touchette}},\ }\bibfield  {title} {\bibinfo {title} {Nonequilibrium microcanonical and canonical ensembles and their equivalence},\ }\href {https://doi.org/10.1103/PhysRevLett.111.120601} {\bibfield  {journal} {\bibinfo  {journal} {Phys. Rev. Lett.}\ }\textbf {\bibinfo {volume} {111}},\ \bibinfo {pages} {120601} (\bibinfo {year} {2013})}\BibitemShut {NoStop}%
\bibitem [{\citenamefont {Garrahan}\ \emph {et~al.}(2009)\citenamefont {Garrahan}, \citenamefont {Jack}, \citenamefont {Lecomte}, \citenamefont {Pitard}, \citenamefont {van Duijvendijk},\ and\ \citenamefont {van Wijland}}]{garrahan2009first}%
  \BibitemOpen
  \bibfield  {author} {\bibinfo {author} {\bibfnamefont {J.~P.}\ \bibnamefont {Garrahan}}, \bibinfo {author} {\bibfnamefont {R.~L.}\ \bibnamefont {Jack}}, \bibinfo {author} {\bibfnamefont {V.}~\bibnamefont {Lecomte}}, \bibinfo {author} {\bibfnamefont {E.}~\bibnamefont {Pitard}}, \bibinfo {author} {\bibfnamefont {K.}~\bibnamefont {van Duijvendijk}},\ and\ \bibinfo {author} {\bibfnamefont {F.}~\bibnamefont {van Wijland}},\ }\bibfield  {title} {\bibinfo {title} {First-order dynamical phase transition in models of glasses: an approach based on ensembles of histories},\ }\href {https://doi.org/10.1088/1751-8113/42/7/075007} {\bibfield  {journal} {\bibinfo  {journal} {J. Phys. A}\ }\textbf {\bibinfo {volume} {42}},\ \bibinfo {pages} {075007} (\bibinfo {year} {2009})}\BibitemShut {NoStop}%
\bibitem [{\citenamefont {Jack}\ and\ \citenamefont {Sollich}(2010)}]{Jack2010}%
  \BibitemOpen
  \bibfield  {author} {\bibinfo {author} {\bibfnamefont {R.~L.}\ \bibnamefont {Jack}}\ and\ \bibinfo {author} {\bibfnamefont {P.}~\bibnamefont {Sollich}},\ }\bibfield  {title} {\bibinfo {title} {Large deviations and ensembles of trajectories in stochastic models},\ }\href {https://doi.org/10.1143/PTPS.184.304} {\bibfield  {journal} {\bibinfo  {journal} {Prog. Theor. Phys. Suppl.}\ }\textbf {\bibinfo {volume} {184}},\ \bibinfo {pages} {304} (\bibinfo {year} {2010})}\BibitemShut {NoStop}%
\bibitem [{\citenamefont {Nyawo}\ and\ \citenamefont {Touchette}(2017)}]{Nyawo2017}%
  \BibitemOpen
  \bibfield  {author} {\bibinfo {author} {\bibfnamefont {P.~T.}\ \bibnamefont {Nyawo}}\ and\ \bibinfo {author} {\bibfnamefont {H.}~\bibnamefont {Touchette}},\ }\bibfield  {title} {\bibinfo {title} {A minimal model of dynamical phase transition},\ }\href {https://doi.org/10.1209/0295-5075/116/50009} {\bibfield  {journal} {\bibinfo  {journal} {EPL}\ }\textbf {\bibinfo {volume} {116}},\ \bibinfo {pages} {50009} (\bibinfo {year} {2017})}\BibitemShut {NoStop}%
\bibitem [{\citenamefont {Nyawo}\ and\ \citenamefont {Touchette}(2018)}]{Nyawo2018}%
  \BibitemOpen
  \bibfield  {author} {\bibinfo {author} {\bibfnamefont {P.~T.}\ \bibnamefont {Nyawo}}\ and\ \bibinfo {author} {\bibfnamefont {H.}~\bibnamefont {Touchette}},\ }\bibfield  {title} {\bibinfo {title} {Dynamical phase transition in drifted brownian motion},\ }\href {https://doi.org/10.1103/PhysRevE.98.052103} {\bibfield  {journal} {\bibinfo  {journal} {Phys. Rev. E}\ }\textbf {\bibinfo {volume} {98}},\ \bibinfo {pages} {052103} (\bibinfo {year} {2018})}\BibitemShut {NoStop}%
\bibitem [{\citenamefont {Hatano}\ and\ \citenamefont {Nelson}(1997)}]{Hatano1997}%
  \BibitemOpen
  \bibfield  {author} {\bibinfo {author} {\bibfnamefont {N.}~\bibnamefont {Hatano}}\ and\ \bibinfo {author} {\bibfnamefont {D.~R.}\ \bibnamefont {Nelson}},\ }\bibfield  {title} {\bibinfo {title} {Vortex pinning and non-{H}ermitian quantum mechanics},\ }\href {https://doi.org/10.1103/PhysRevB.56.8651} {\bibfield  {journal} {\bibinfo  {journal} {Phys. Rev. B}\ }\textbf {\bibinfo {volume} {56}},\ \bibinfo {pages} {8651} (\bibinfo {year} {1997})}\BibitemShut {NoStop}%
\bibitem [{\citenamefont {Nieto}(2002)}]{Nieto2002}%
  \BibitemOpen
  \bibfield  {author} {\bibinfo {author} {\bibfnamefont {M.~M.}\ \bibnamefont {Nieto}},\ }\bibfield  {title} {\bibinfo {title} {Existence of bound states in continuous {$0<D<\infty$} dimensions},\ }\href {https://doi.org/10.1016/S0375-9601(01)00827-1} {\bibfield  {journal} {\bibinfo  {journal} {Phys. Lett. A}\ }\textbf {\bibinfo {volume} {293}},\ \bibinfo {pages} {10} (\bibinfo {year} {2002})}\BibitemShut {NoStop}%
\bibitem [{\citenamefont {Yamamoto}(1996)}]{Yamamoto1996}%
  \BibitemOpen
  \bibfield  {author} {\bibinfo {author} {\bibfnamefont {A.}~\bibnamefont {Yamamoto}},\ }\bibfield  {title} {\bibinfo {title} {Crystallography of quasiperiodic crystals},\ }\href {https://doi.org/10.1107/S0108767396000967} {\bibfield  {journal} {\bibinfo  {journal} {Acta Crystallogr. Sect. A}\ }\textbf {\bibinfo {volume} {52}},\ \bibinfo {pages} {509} (\bibinfo {year} {1996})}\BibitemShut {NoStop}%
\bibitem [{\citenamefont {Nagai}\ \emph {et~al.}(2024)\citenamefont {Nagai}, \citenamefont {Iwasaki}, \citenamefont {Kitahara}, \citenamefont {Takagiwa}, \citenamefont {Kimura},\ and\ \citenamefont {Shiga}}]{Nagai2024}%
  \BibitemOpen
  \bibfield  {author} {\bibinfo {author} {\bibfnamefont {Y.}~\bibnamefont {Nagai}}, \bibinfo {author} {\bibfnamefont {Y.}~\bibnamefont {Iwasaki}}, \bibinfo {author} {\bibfnamefont {K.}~\bibnamefont {Kitahara}}, \bibinfo {author} {\bibfnamefont {Y.}~\bibnamefont {Takagiwa}}, \bibinfo {author} {\bibfnamefont {K.}~\bibnamefont {Kimura}},\ and\ \bibinfo {author} {\bibfnamefont {M.}~\bibnamefont {Shiga}},\ }\bibfield  {title} {\bibinfo {title} {High-temperature atomic diffusion and specific heat in quasicrystals},\ }\href {https://doi.org/10.1103/physrevlett.132.196301} {\bibfield  {journal} {\bibinfo  {journal} {Phys. Rev. Lett.}\ }\textbf {\bibinfo {volume} {132}},\ \bibinfo {pages} {196301} (\bibinfo {year} {2024})}\BibitemShut {NoStop}%
\bibitem [{\citenamefont {Randeria}\ and\ \citenamefont {Taylor}(2014)}]{Randeria2014}%
  \BibitemOpen
  \bibfield  {author} {\bibinfo {author} {\bibfnamefont {M.}~\bibnamefont {Randeria}}\ and\ \bibinfo {author} {\bibfnamefont {E.}~\bibnamefont {Taylor}},\ }\bibfield  {title} {\bibinfo {title} {Crossover from {B}ardeen-{C}ooper-{S}chrieffer to {B}ose-{E}instein condensation and the unitary {F}ermi gas},\ }\href {https://doi.org/10.1146/annurev-conmatphys-031113-133829} {\bibfield  {journal} {\bibinfo  {journal} {Annu. Rev. Condens. Matter Phys.}\ }\textbf {\bibinfo {volume} {5}},\ \bibinfo {pages} {209} (\bibinfo {year} {2014})}\BibitemShut {NoStop}%
\bibitem [{\citenamefont {Kanazawa}\ \emph {et~al.}()\citenamefont {Kanazawa}, \citenamefont {Kawaguchi},\ and\ \citenamefont {Adachi}}]{Full_arxiv}%
  \BibitemOpen
  \bibfield  {author} {\bibinfo {author} {\bibfnamefont {T.}~\bibnamefont {Kanazawa}}, \bibinfo {author} {\bibfnamefont {K.}~\bibnamefont {Kawaguchi}},\ and\ \bibinfo {author} {\bibfnamefont {K.}~\bibnamefont {Adachi}},\ }\href {http://arxiv.org/abs/2407.14090} {\bibinfo {title} {{Universality in the dynamical phase transitions of Brownian motion}}},\ \Eprint {https://arxiv.org/abs/2407.14090} {arXiv:2407.14090} \BibitemShut {NoStop}%
\bibitem [{\citenamefont {Touchette}(2018)}]{Touchette2018}%
  \BibitemOpen
  \bibfield  {author} {\bibinfo {author} {\bibfnamefont {H.}~\bibnamefont {Touchette}},\ }\bibfield  {title} {\bibinfo {title} {Introduction to dynamical large deviations of {M}arkov processes},\ }\href {https://doi.org/10.1016/j.physa.2017.10.046} {\bibfield  {journal} {\bibinfo  {journal} {Physica A}\ }\textbf {\bibinfo {volume} {504}},\ \bibinfo {pages} {5} (\bibinfo {year} {2018})}\BibitemShut {NoStop}%
\bibitem [{\citenamefont {Chetrite}\ and\ \citenamefont {Touchette}(2015)}]{Chetrite2015}%
  \BibitemOpen
  \bibfield  {author} {\bibinfo {author} {\bibfnamefont {R.}~\bibnamefont {Chetrite}}\ and\ \bibinfo {author} {\bibfnamefont {H.}~\bibnamefont {Touchette}},\ }\bibfield  {title} {\bibinfo {title} {Nonequilibrium {M}arkov processes conditioned on large deviations},\ }\href {https://doi.org/10.1007/s00023-014-0375-8} {\bibfield  {journal} {\bibinfo  {journal} {Ann. Henri Poincar{\'e}}\ }\textbf {\bibinfo {volume} {16}},\ \bibinfo {pages} {2005} (\bibinfo {year} {2015})}\BibitemShut {NoStop}%
\bibitem [{\citenamefont {Agranov}\ \emph {et~al.}(2020)\citenamefont {Agranov}, \citenamefont {Zilber}, \citenamefont {Smith}, \citenamefont {Admon}, \citenamefont {Roichman},\ and\ \citenamefont {Meerson}}]{Agranov2020}%
  \BibitemOpen
  \bibfield  {author} {\bibinfo {author} {\bibfnamefont {T.}~\bibnamefont {Agranov}}, \bibinfo {author} {\bibfnamefont {P.}~\bibnamefont {Zilber}}, \bibinfo {author} {\bibfnamefont {N.~R.}\ \bibnamefont {Smith}}, \bibinfo {author} {\bibfnamefont {T.}~\bibnamefont {Admon}}, \bibinfo {author} {\bibfnamefont {Y.}~\bibnamefont {Roichman}},\ and\ \bibinfo {author} {\bibfnamefont {B.}~\bibnamefont {Meerson}},\ }\bibfield  {title} {\bibinfo {title} {Airy distribution: Experiment, large deviations, and additional statistics},\ }\href {https://doi.org/10.1103/PhysRevResearch.2.013174} {\bibfield  {journal} {\bibinfo  {journal} {Phys. Rev. Res.}\ }\textbf {\bibinfo {volume} {2}},\ \bibinfo {pages} {013174} (\bibinfo {year} {2020})}\BibitemShut {NoStop}%
\bibitem [{\citenamefont {Landau}\ and\ \citenamefont {Lifshitz}(1981)}]{Landau_quantum}%
  \BibitemOpen
  \bibfield  {author} {\bibinfo {author} {\bibfnamefont {L.~D.}\ \bibnamefont {Landau}}\ and\ \bibinfo {author} {\bibfnamefont {E.~M.}\ \bibnamefont {Lifshitz}},\ }\href@noop {} {\emph {\bibinfo {title} {Quantum Mechanics: Nonrelativistic Theory}}},\ \bibinfo {series} {Course of Theoretical Physics}, Vol.~\bibinfo {volume} {3}\ (\bibinfo  {publisher} {Butterworth-Heinemann},\ \bibinfo {year} {1981})\BibitemShut {NoStop}%
\bibitem [{\citenamefont {Sahu}\ \emph {et~al.}(1989)\citenamefont {Sahu}, \citenamefont {Khan}, \citenamefont {Shastry}, \citenamefont {Dey},\ and\ \citenamefont {Phatak}}]{Sahu1989}%
  \BibitemOpen
  \bibfield  {author} {\bibinfo {author} {\bibfnamefont {B.}~\bibnamefont {Sahu}}, \bibinfo {author} {\bibfnamefont {M.~Z.~R.}\ \bibnamefont {Khan}}, \bibinfo {author} {\bibfnamefont {C.~S.}\ \bibnamefont {Shastry}}, \bibinfo {author} {\bibfnamefont {B.}~\bibnamefont {Dey}},\ and\ \bibinfo {author} {\bibfnamefont {S.~C.}\ \bibnamefont {Phatak}},\ }\bibfield  {title} {\bibinfo {title} {Interesting features of relativistic and nonrelativistic quantal bound states in one, two, and three dimensions},\ }\href {https://doi.org/10.1119/1.15841} {\bibfield  {journal} {\bibinfo  {journal} {Am. J. Phys.}\ }\textbf {\bibinfo {volume} {57}},\ \bibinfo {pages} {886} (\bibinfo {year} {1989})}\BibitemShut {NoStop}%
\bibitem [{\citenamefont {Landau}\ and\ \citenamefont {Lifshitz}(1980)}]{Landau_statphys}%
  \BibitemOpen
  \bibfield  {author} {\bibinfo {author} {\bibfnamefont {L.~D.}\ \bibnamefont {Landau}}\ and\ \bibinfo {author} {\bibfnamefont {E.~M.}\ \bibnamefont {Lifshitz}},\ }\href@noop {} {\emph {\bibinfo {title} {Statistical Physics}}},\ \bibinfo {series} {Course of Theoretical Physics}, Vol.~\bibinfo {volume} {5}\ (\bibinfo  {publisher} {Butterworth-Heinemann},\ \bibinfo {year} {1980})\BibitemShut {NoStop}%
\bibitem [{\citenamefont {Rubinstein}\ and\ \citenamefont {Colby}(2003)}]{Rubinstein2003}%
  \BibitemOpen
  \bibfield  {author} {\bibinfo {author} {\bibfnamefont {M.}~\bibnamefont {Rubinstein}}\ and\ \bibinfo {author} {\bibfnamefont {R.~H.}\ \bibnamefont {Colby}},\ }\href@noop {} {\emph {\bibinfo {title} {Polymer Physics}}}\ (\bibinfo  {publisher} {Oxford University Press},\ \bibinfo {year} {2003})\BibitemShut {NoStop}%
\bibitem [{\citenamefont {Hohenberg}\ and\ \citenamefont {Halperin}(1977)}]{Hohenberg1977}%
  \BibitemOpen
  \bibfield  {author} {\bibinfo {author} {\bibfnamefont {P.~C.}\ \bibnamefont {Hohenberg}}\ and\ \bibinfo {author} {\bibfnamefont {B.~I.}\ \bibnamefont {Halperin}},\ }\bibfield  {title} {\bibinfo {title} {Theory of dynamic critical phenomena},\ }\href {https://doi.org/10.1103/RevModPhys.49.435} {\bibfield  {journal} {\bibinfo  {journal} {Rev. Mod. Phys.}\ }\textbf {\bibinfo {volume} {49}},\ \bibinfo {pages} {435} (\bibinfo {year} {1977})}\BibitemShut {NoStop}%
\bibitem [{\citenamefont {Chaikin}\ and\ \citenamefont {Lubensky}(1995)}]{Chaikin1995}%
  \BibitemOpen
  \bibfield  {author} {\bibinfo {author} {\bibfnamefont {P.~M.}\ \bibnamefont {Chaikin}}\ and\ \bibinfo {author} {\bibfnamefont {T.~C.}\ \bibnamefont {Lubensky}},\ }\href@noop {} {\emph {\bibinfo {title} {Principles of Condensed Matter Physics}}}\ (\bibinfo  {publisher} {Cambridge University Press},\ \bibinfo {year} {1995})\BibitemShut {NoStop}%
\bibitem [{\citenamefont {Klaus}(1977)}]{Klaus1977}%
  \BibitemOpen
  \bibfield  {author} {\bibinfo {author} {\bibfnamefont {M.}~\bibnamefont {Klaus}},\ }\bibfield  {title} {\bibinfo {title} {On the bound state of {S}chr{\"o}dinger operators in one dimension},\ }\href {https://doi.org/10.1016/0003-4916(77)90015-X} {\bibfield  {journal} {\bibinfo  {journal} {Ann. Phys. (N. Y.)}\ }\textbf {\bibinfo {volume} {108}},\ \bibinfo {pages} {288} (\bibinfo {year} {1977})}\BibitemShut {NoStop}%
\bibitem [{\citenamefont {Klaus}\ and\ \citenamefont {Simon}(1980)}]{Klaus1980}%
  \BibitemOpen
  \bibfield  {author} {\bibinfo {author} {\bibfnamefont {M.}~\bibnamefont {Klaus}}\ and\ \bibinfo {author} {\bibfnamefont {B.}~\bibnamefont {Simon}},\ }\bibfield  {title} {\bibinfo {title} {Coupling constant thresholds in nonrelativistic quantum mechanics. {I}. short-range two-body case},\ }\href {https://doi.org/10.1016/0003-4916(80)90338-3} {\bibfield  {journal} {\bibinfo  {journal} {Ann. Phys. (N. Y.)}\ }\textbf {\bibinfo {volume} {130}},\ \bibinfo {pages} {251} (\bibinfo {year} {1980})}\BibitemShut {NoStop}%
\bibitem [{\citenamefont {Simon}(1976)}]{Simon1976}%
  \BibitemOpen
  \bibfield  {author} {\bibinfo {author} {\bibfnamefont {B.}~\bibnamefont {Simon}},\ }\bibfield  {title} {\bibinfo {title} {The bound state of weakly coupled {S}chr{\"o}dinger operators in one and two dimensions},\ }\href {https://doi.org/10.1016/0003-4916(76)90038-5} {\bibfield  {journal} {\bibinfo  {journal} {Ann. Phys. (N. Y.)}\ }\textbf {\bibinfo {volume} {97}},\ \bibinfo {pages} {279} (\bibinfo {year} {1976})}\BibitemShut {NoStop}%
\bibitem [{\citenamefont {Bouchaud}\ and\ \citenamefont {Potters}(2003)}]{Bouchaud2003}%
  \BibitemOpen
  \bibfield  {author} {\bibinfo {author} {\bibfnamefont {J.-P.}\ \bibnamefont {Bouchaud}}\ and\ \bibinfo {author} {\bibfnamefont {M.}~\bibnamefont {Potters}},\ }\href@noop {} {\emph {\bibinfo {title} {Theory of financial risk and derivative pricing: from statistical physics to risk management}}}\ (\bibinfo  {publisher} {Cambridge university press},\ \bibinfo {year} {2003})\BibitemShut {NoStop}%
\bibitem [{\citenamefont {Sohl-Dickstein}\ \emph {et~al.}(2015)\citenamefont {Sohl-Dickstein}, \citenamefont {Weiss}, \citenamefont {Maheswaranathan},\ and\ \citenamefont {Ganguli}}]{Sohl2015}%
  \BibitemOpen
  \bibfield  {author} {\bibinfo {author} {\bibfnamefont {J.}~\bibnamefont {Sohl-Dickstein}}, \bibinfo {author} {\bibfnamefont {E.}~\bibnamefont {Weiss}}, \bibinfo {author} {\bibfnamefont {N.}~\bibnamefont {Maheswaranathan}},\ and\ \bibinfo {author} {\bibfnamefont {S.}~\bibnamefont {Ganguli}},\ }\bibfield  {title} {\bibinfo {title} {Deep unsupervised learning using nonequilibrium thermodynamics},\ }in\ \href {https://proceedings.mlr.press/v37/sohl-dickstein15.html} {\emph {\bibinfo {booktitle} {International Conference on Machine Learning}}}\ (\bibinfo  {publisher} {PMLR},\ \bibinfo {year} {2015})\ p.\ \bibinfo {pages} {2256}\BibitemShut {NoStop}%
\end{thebibliography}
\end{document}